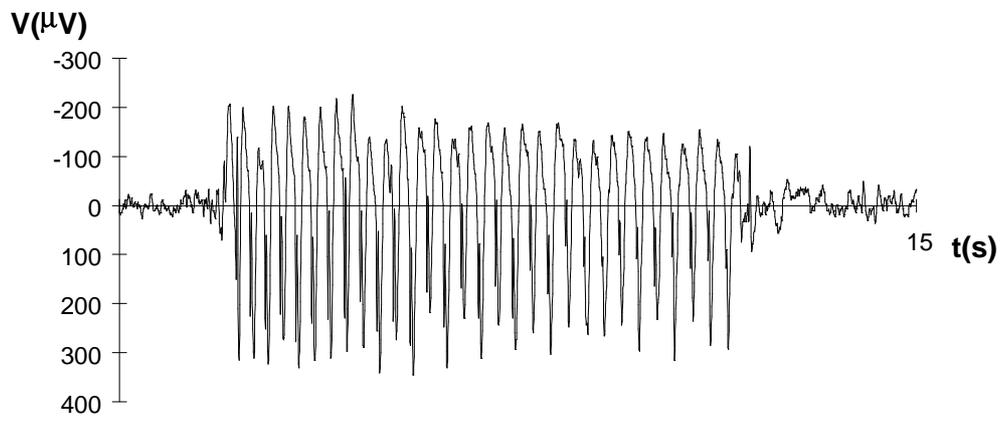

FIGURE 1 a)

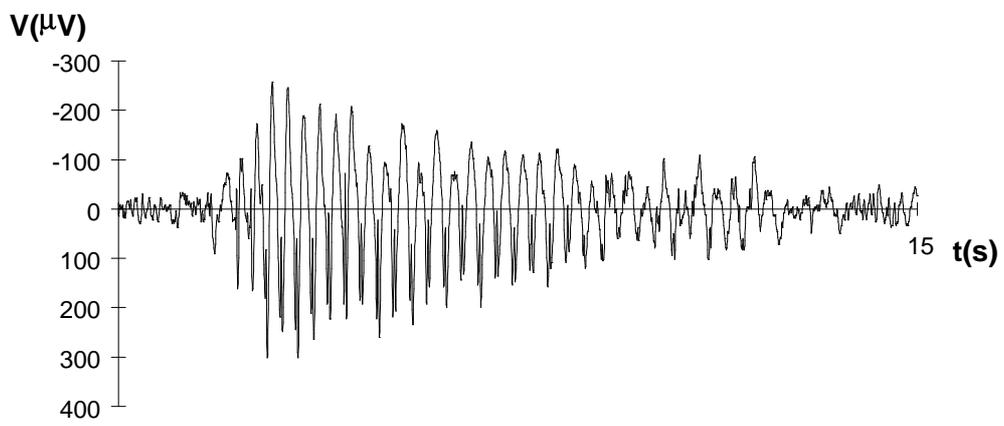

FIGURE 1 b)

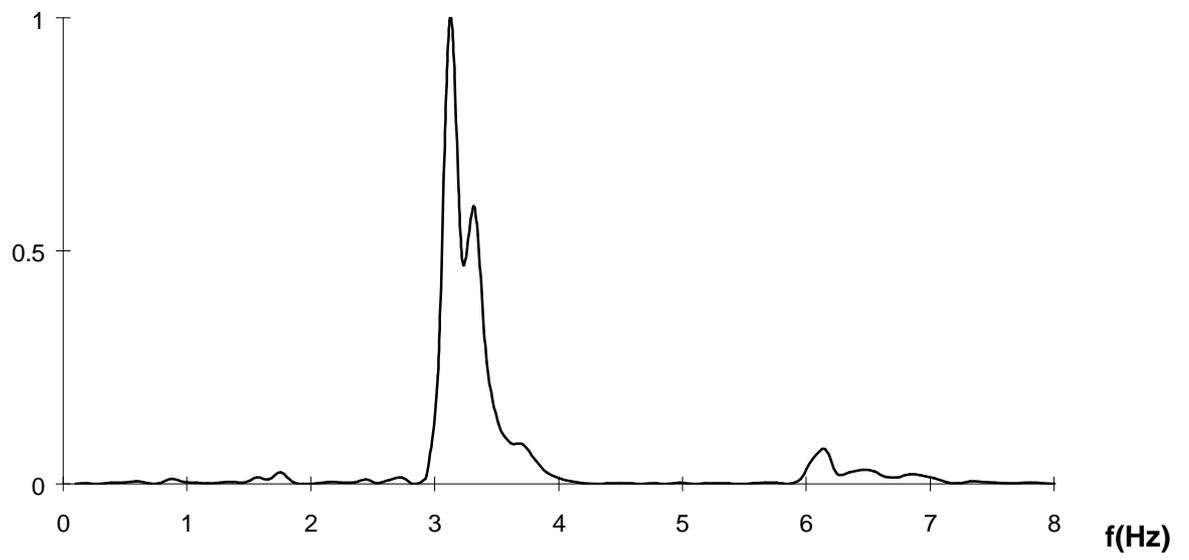

FIGURE 2 a)

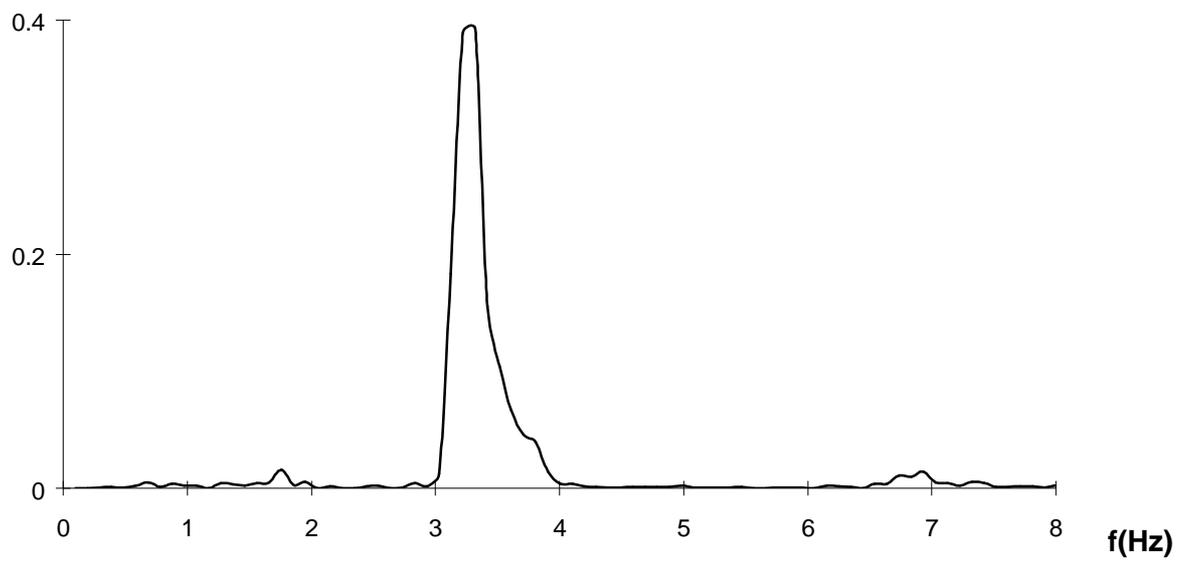

FIGURE 2 b)

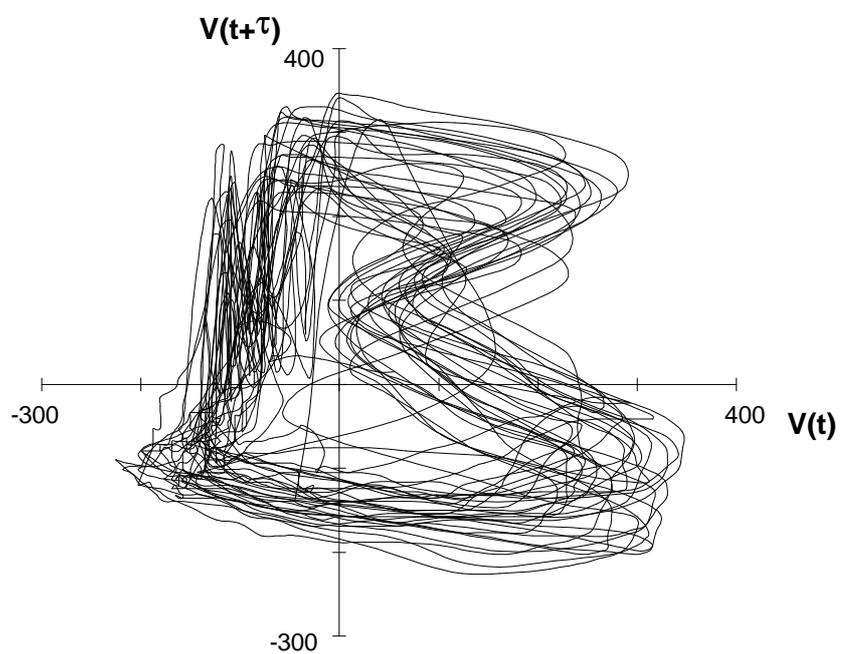

FIGURE 3 a)

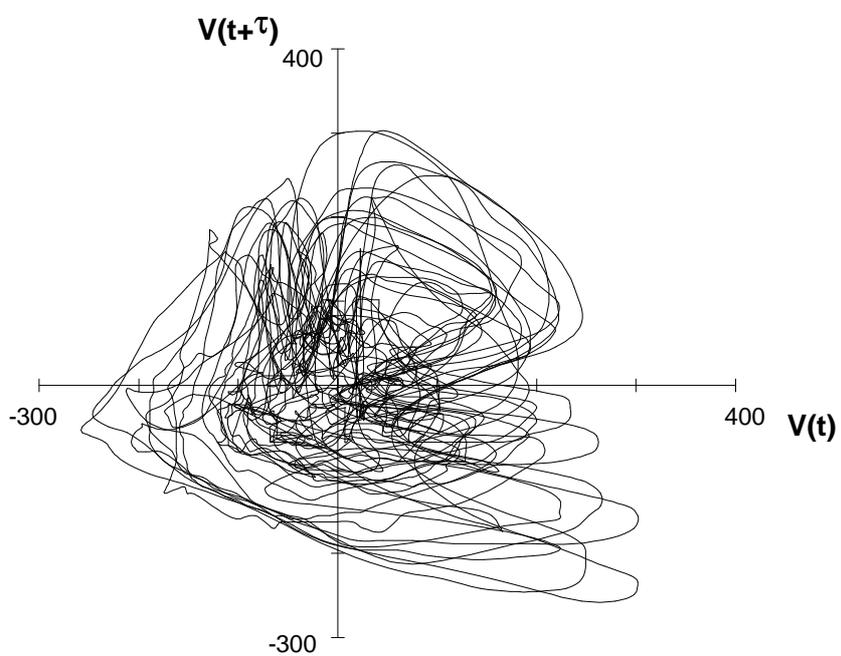

FIGURE 3 b)

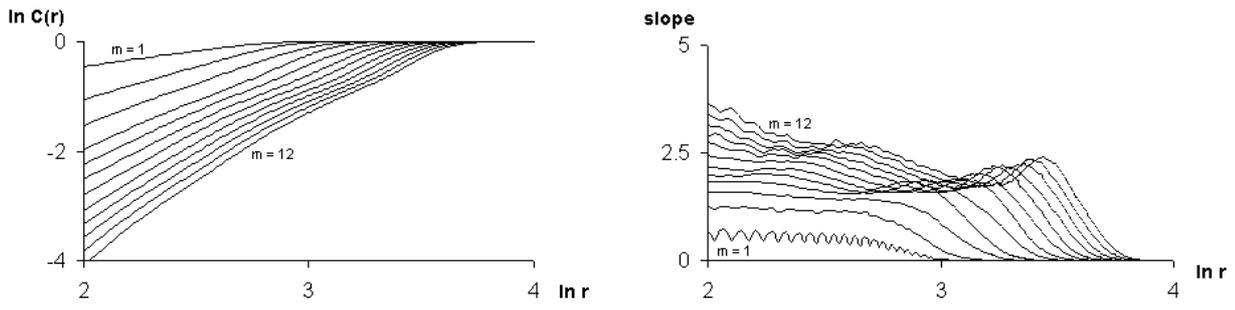

FIGURE 4 a)

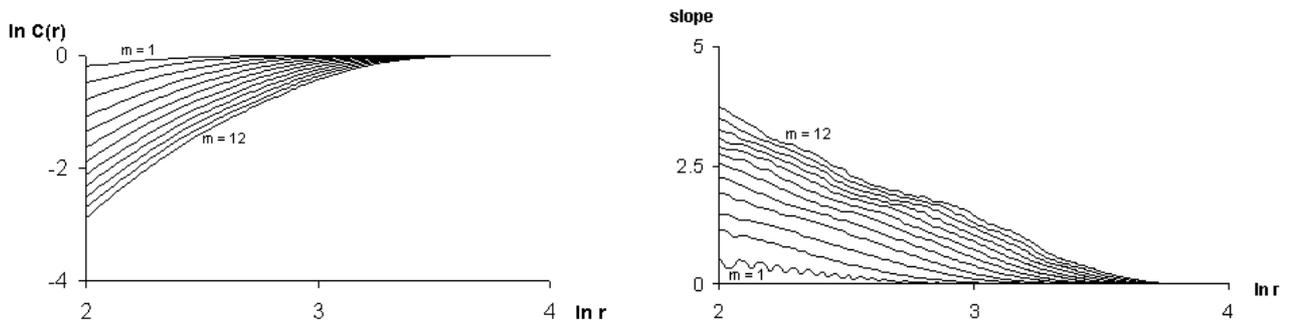

FIGURE 4 b)

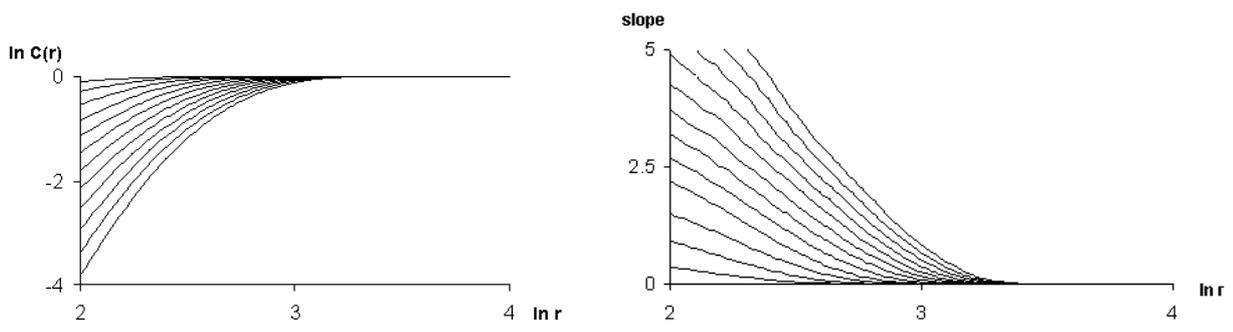

FIGURE 5

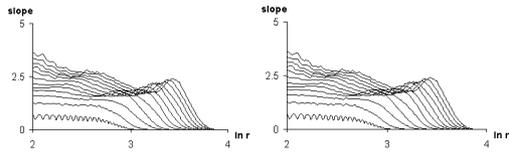

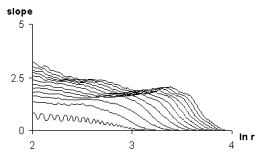
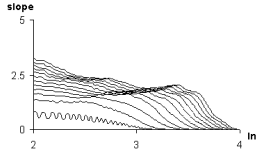
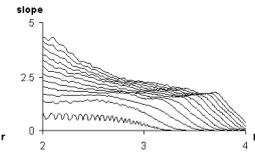
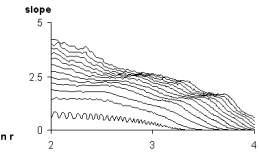
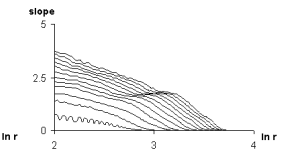

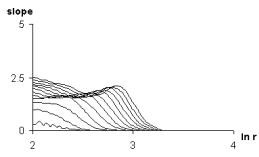
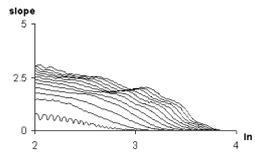
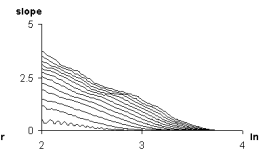
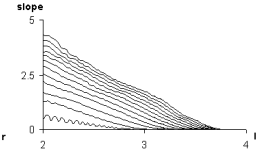
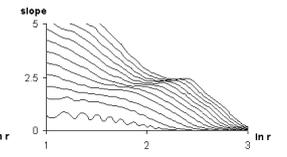

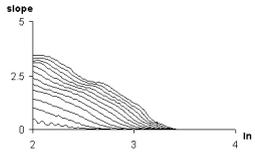
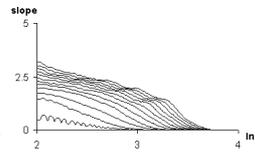
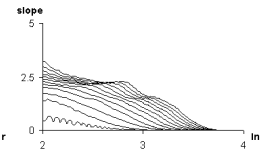
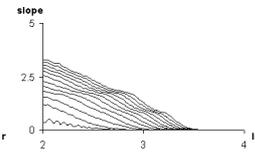
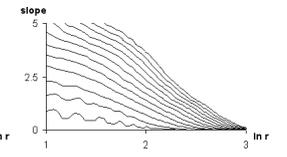

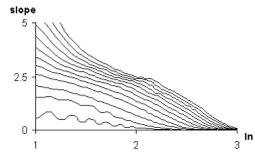
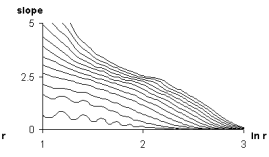
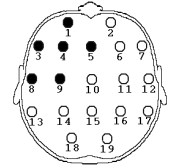

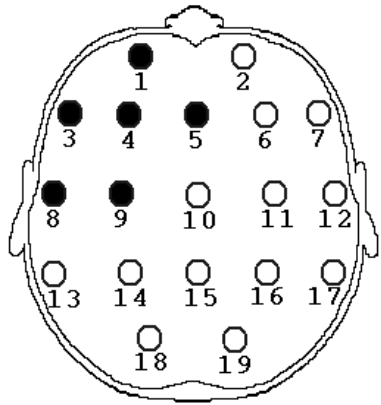

FIGURE 7 a)

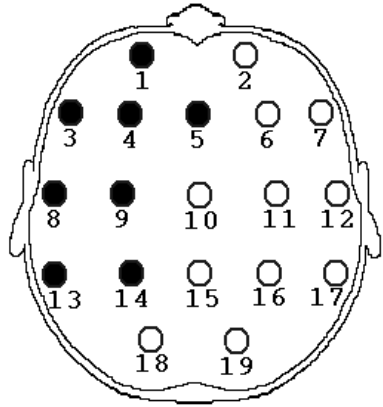

FIGURE 7 b)

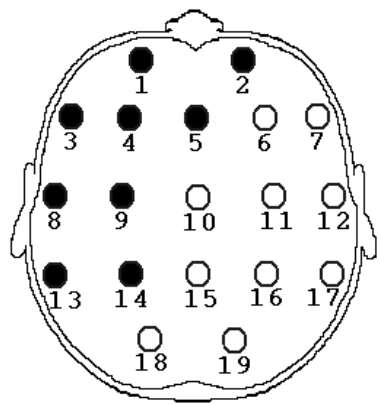

FIGURE 7 c)

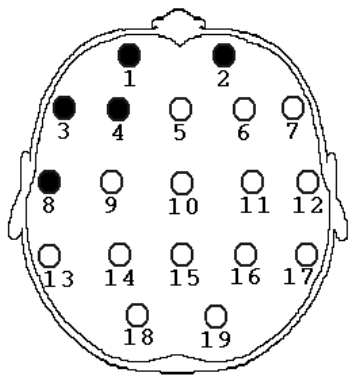

FIGURE 8 a)

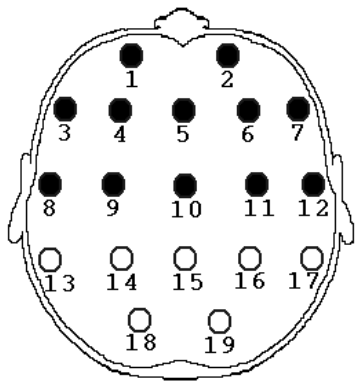

FIGURE 8 b)

# Correlation Dimension Maps of EEG

# from epileptic Absences

[*]Carla Silva (MSc), [+]Iveta R. Pimentel (DPhil), [*]Alexandre Andrade (MSc),

[#]John P. Foreid (MD) and [*]Eduardo Ducla-Soares (PhD)

[*] Institute of Biophysics and Biomedical Engineering and [+] Condensed Matter Physics Centre, University of Lisbon, Campo Grande, 1700 Lisboa. [#] Portuguese Institute of Oncology, Lisboa.

**Summary**

Purpose: The understanding of brain activity, and in particular events such as epileptic seizures, lies on the characterisation of the dynamics of the neural networks. The theory of non-linear dynamics provides signal analysis techniques which may give new information on the behaviour of such networks. Methods: We calculated correlation dimension maps for 19-channel EEG data from 3 patients with a total of 7 absence seizures. The signals were analysed before, during and after the seizures. Phase randomised surrogate data was used to test chaos. Results: In the seizures of two patients we could distinguish two dynamical regions on the cerebral cortex, one that seemed to exhibit chaos whereas the other seemed to exhibit noise. The pattern shown is essentially the same for seizures triggered by hyperventilation, but differ for seizures triggered by light flashes. The chaotic dynamics that one seems to observe is determined by a small number of variables and has low complexity. On the other hand, in the seizures of another patient no chaotic region was found. Before and during the seizures no chaos was found either, in all cases. Conclusions: The application of non-linear signal analysis revealed the existence of differences in the spatial dynamics associated to absence seizures. This may contribute to the understanding of those seizures and be of assistance in clinical diagnosis.

## 1. Introduction

Over the last few years there has been an increasing interest in the application of non-linear dynamics theory, commonly referred as chaos theory, to brain activity. Those studies have mostly been concerned with EEG signals from intracranial or scalp recordings, in animals or human subjects, and considered in particular epilepsy, sleep, cognitive and evoked responses (see e.g. Elbert et al 1994, and Pritchards and Duke 1995).

Chaos theory (Schuster 1984, Bassingthwaighte et al 1994, Elbert et al. 1994) allows a characterisation of the dynamics of complex systems from the analysis of a signal generated by the system, which consists of a series of measurements in time of a pertinent and easily accessible variable. In brain studies one uses EEG data to investigate the dynamics of the brain neuronal networks.

EEG signals show in general great irregularity that may have different origins, i.e. it may be simply due to noise or otherwise may reflect the presence of chaos. Chaos is irregular behaviour that occurs in deterministic systems with a small number of independent variables that are non-linear. Noise is simply produced by random fluctuation of many variables. Chaos theory allows a distinction on wether the irregularities in the EEG signal are due to chaos or noise. Such approach may therefore provide a new insight into the dynamics of brain activity since the two situations involve different underlying mechanisms. In the presence of chaos, the complexity of the dynamics can be quantified in terms of the properties of the attractor in phase-space, e.g. its correlation dimension $D_2$. Dimensional analysis may therefore provide a classification of brain activity in terms of its complexity. However a careful discussion is necessary to distinguish chaos from noise because finding a correlation dimension $D_2$ is a necessary but not a sufficient condition for chaos. Even though the presence of chaos cannot be assured, the correlation dimension analysis may provide valuable information as a tool to detect differences in the dynamic behaviour associated to different degrees of determinism.

There are different types of epilepsy (Lopes da Silva and Niedermeyer 1993), with a focal or generalised nature. Epileptic seizures may occur spontaneously or may be induced by various means. Well-controlled intracranial EEG recordings were performed in rats with focal epilepsy and the data was analysed using chaos theory (Pijn et al. 1991), in order to test the ability of this tool to detect the epileptogenic focus and the spread of the seizure activity. A decrease of the correlation dimension $D_2$ was observed at the seizure onset. Later, chaos analysis was applied to intracranial EEG recordings from a group of patients with unilateral temporal lobe epilepsy (Lehnerty and Elger 1995). A low value of $D_2$ was found during the seizure, again especially at the zone of ictal onset. Recently, Pijn et al. (1997) carried out a very thorough analysis of chaos in intracranially EEG recorded signals, during interictal and ictal states, of temporal lobe epileptic patients. Although they were not able to ascribe a value of $D_2$ to the signals showing seizure activity, they found significant evidence of the existence of a considerable degree of determinism in the system generating those signals. On the contrary, they found that signals from areas that did not show seizure activity were almost indistinguishable from their randomised versions, regardless of whether they were recorded during interictal or ictal states. They then concluded that, in the particular case studied, chaos analysis yields good results in terms of locating the epileptogenic region and following the ictal spread throughout the brain. Looking at a generalised type of epilepsy, Babloyanz and Destexhe (1986), who were the first to apply chaos theory to epileptic activity, analysed EEG signals from an absence seizure, recorded at two different sites on the brain. They found chaos in the signals, with a low value of $D_2$ which implies low complexity. Chaotic behaviour with a higher value of $D_2$ was also reported (Frank et al. 1990) for epileptic activity but, in this case, the seizures were a generalised state with both absence and grand mal events.

In this work we present a chaos analysis of 19-channel EEG data normally recorded in a clinical setting, from patients with absence seizures. This type of generalised epilepsy usually invades the entire cerebral cortex and shows, in general, a bilateral symmetry between the two hemispheres. Our purpose on analysing such a set of channels is to detect possible spatial variations in the cerebral dynamics. We present maps of the correlation dimension $D_2$ over the brain, for a total of 7 seizures from 3 patients, stimulated either by hyperventilation or light flashes. Those seizures were selected from a more extended set of data (18 seizures from 9 patients) as they are representative of the different dynamical behaviours that we found and that will be presented in this work. We analyse the EEG signals before, during and after the epileptic seizures, in order to look at differences in the dynamics of the neuronal networks at the various states. This may allow a characterisation of the epileptic activity. In order to have a stronger test to distinguish chaos from noise we compare our original EEG signals with surrogate data obtained through phase randomisation of their Fourier components (Pijn et al. 1991).

## 2. Materials and Methods

### 2.1 Data acquistion

Electroencephalographic recordings of 19 channels in a standard 10/20 referential configuration, were taken from patients with absences. The signals were recorded before, during and after the epileptic seizures. These were triggered by hyperventilation or light flashes. The reference potential was given by the average of the signals at electrodes located on each side of the chin. A Bio-Logic recording system was used with an acquisition rate of 100Hz or 200Hz, and the signals were filtered high-pass 1Hz and for the sampling rate of 200Hz also with a low-pass 70Hz (the highest value available for spontaneous activity recording, in the system used).

## 2.2 Data analysis

We start by presenting the most traditional forms of data analysis, direct inspection of the time series, power spectrum and auto-correlation, and then evolve into techniques of chaos theory, which will allow a distinction on wether the irregularities in the EEG signals are noise or possibly chaos; in the latter case a quantitative characterisation of the complexity of the dynamics is provided in terms of the correlation dimension $D_2$.

*EEG signal.* Signals associated to a triggered epileptic seizure were observed over the entire scalp. Fig. 1 shows EEG recordings at two different channels on a patient. During the seizure the electrical potential of the brain V(t) suddenly increases by typically a factor of ten, switching into a series of spike - slow wave complexes with a dominant frequency of $\approx$ 3Hz, which shows some irregularity. All the signals observed, at the different channels on each patient, seem to exhibit the same structure, though with differences in amplitude depending on the site. The measured seizures had durations from about 7 to 15s.

*Power spectrum.* The power spectrum gives information on the frequency values present in the signal and their weights. Although the power spectrum has played a major role in data analysis, it misses crucial information on the phase contents of the signals. Fig. 2 shows the power spectrum of the signals in Fig. 1. One can see a large peak at $\approx$ 3Hz (and the harmonic at $\approx$ 6Hz), and a broad band around that value extending from lower to higher frequencies. This band is associated to the irregularities in the EEG signal. The power spectrum for all the signals, at the different channels on a patient, seem to exhibit the same structure, though again with differences in amplitude depending on the site.

*Auto-correlation.* The auto-correlation function, defined as $A(t) = \int V(t')V(\tau + t')dt'$, gives information on the correlations in time present in the signal. We calculated the auto-correlation for the EEG signal at the different channels in each patient and found that it is an oscillating function with a decaying envelope. This indicates that the signal has important correlations but they decay which implies that there is loss of information in time.

The scalp electrical potential of the brain that one measures results from a set of $d$ independent variables $X_i(t)$, which specify the state of the system at any time. Those variables define a d-dimensional phase-space, in which the state of the system is represented by $\vec{X}(t) = [X_1, X_2, \ldots, X_d]$. A chaotic system is characterised by having an attractor, that is a limiting set of points to which all trajectories are attracted in the phase-space. The attractor has a fractal structure and the complexity of the dynamics can be quantified in terms of the correlation dimension of the attractor $D_2$. $D_2$ is a lower estimator of the Hausdorff dimension D which measures the occupation of the attractor in phase-space ($D_2 < D < d$).

In order to investigate the chaotic nature of the dynamics, and possibly measure its complexity, one needs to reconstruct the dynamics of the system in phase-space from the measured time series V(t). This can be done by the "method of time delays" (Takens 1981), which is based on building vectors associated to each time $t_i$ on the time series, with components that are the signal at time $t_i$ plus an increasing number of time delays $\tau$ :

$$\vec{V}_m(t_i) = \left[ V(t_i), V(t_i + \tau), \dots, V\left(t_i + (m-1)\tau\right) \right]$$

Those vectors create a pseudo phase-space with dimension m, which is topologically equivalent to the original phase space, for m$\geq$ 2d+1. We take $\tau$ equal to the first zero crossing of the auto-correlation, so that the original and delayed signals are not strongly correlated; this is a usual choice for $\tau$ , and the results do not depend significantly on $\tau$ when it is taken within reasonable limits (Schuster 1984, Fraser and Swinney 1986, Libert and Schuster 1989, Bassingthwaighte et al 1994, Elbert et al. 1994).

*Map.* A look into the phase-space of the system can be obtained via the map, that is the plot of V(t+$\tau$ ) *vs* V(t). The map represents a projection of the attractor in the pseudo phase-space and reflects the correlations in the signal. In Fig. 3 we present the maps corresponding to the signals in Fig. 1. The two maps are clearly different: in a) one can identify an attractor which reflects particular correlations in the signal and may therefore imply chaos, whereas in b) the space is more or less uniformly covered which is more characteristic of noise. This kind of analysis is giving for the first time evidence that although the EEG time series look similar in all the channels, indeed different dynamics may be occurring in different areas of the brain.

*Correlation dimension.* Using the Grassberger-Procaccia (1983[a], 1983[b]) algorithm to determine the correlation dimension $D_2$, one defines the correlation integral:

$$C_m(r) = \frac{1}{N^2} \sum_{i \neq j} \theta\left( r - \left| \vec{V}_m(t_i) - \vec{V}_m(t_j) \right| \right)$$

where, $\theta(x) = 1$ if $x \geq 0$, $\theta(x) = 0$ if $x < 0$, and N is the number of points in the time series. $C_m(r)$ measures the fraction of pairs of points in space that are closer than r. If the system is chaotic one has that for sufficiently large m, m>m*, the correlation integral takes the following scaling form, independent of m,

$$C(r) \sim r^{D_2} \quad (1)$$

with the exponent giving the correlation dimension $D_2$ of the attractor corresponding to the measured signal. Hence $D_2$ can be obtained from the slope of ln C(r) *vs* ln r. The quantity m* is the minimal embedding dimension as it is the lowest integer dimension containing the whole attractor; m* gives information on the number of independent variables governing the dynamics of the system.

The correlation integrals associated to the signals in Fig. 1 are shown in Fig. 4. In order to avoid spurious effects we made the "Theiler correction" to auto-correlation in the correlation integrals (Theiler 1986, Theiler 1990). In the plots of the correlation integral, the region that is relevant for the analysis of the epileptic signal is the one corresponding to the larger values of r, the region for the smaller values of r being strongly contaminated by small amplitude noise from different origins. It is clear that the channels in a) and b) exhibit different behaviour. One observes that in a) there is a region where the correlation integral behaves like (1), which means that ln $C_m(r)$ has a constant slope, also independent of m. Such behaviour is translated into the presence of a plateau in the plot of the slopes, and its value gives the correlation dimension $D_2$. The embedding dimension m* is given by the value of m above which the plateau sets in. We therefore conclude that the EEG signal in Fig. 1-a) seems to exhibit chaotic behaviour. By contrast, in b) no plateau is observed in the plot of the slopes, which means that the correlation integral never behaves like (1), and hence the EEG signal in Fig. 1-b) does not exhibit chaotic behaviour. One could expect that even the presence of small amplitude noise could blur the appearance of a plateau. In order to have some sensitivity to that problem, we made a simulation in which we considered a sinusoidal signal mixed with different levels of noise, and verified that even in the presence of noise which is 40% of the signal, the plateau remains defined. This implies that the presence or absence of a plateau in the EEG signals is intrinsic to the signal, up to that level of noise. The results here obtained from the correlation integrals confirm the observations made from the maps.

*Chaos vs noise.* Finding a finite correlation dimension $D_2$ does not however necessarily imply having chaos, because coloured (i.e. power law spectra) noise may also give rise to it (Osborne and Provenzale 1989). In order to distinguish chaos from noise, we built a control signal that has the same power spectrum as the measured signal, but has randomised phases. Then we compared the measured and control signals via the correlation integral slope: only if the two look different may it be concluded that the measured signal is not noise and may be chaotic. Fig. 5 shows the correlation integral and respective slope for the control signal corresponding to the signal of Fig. 1-a), to be compared with Fig 4-a). One clearly sees the difference between the two signals, with an absence of a plateau in the control signal. We therefore conclude that the measured signal seems to exhibit chaotic behaviour. This analysis was done for all the signals showing a finite correlation dimension, and for all we observed that the measured and control signals were different, thus implying a possible presence of chaos.

We recall that, even after passing the phase randomisation test, the existence of a $D_2$ does not necessarily imply chaos, but, as we have just seen, it provides a mean to detect differences in the dynamical behaviour over the scalp.

The application of a correlation dimension analysis implicitly assumes that the signal that we are studying is stationary. An epileptic seizure may naturally have some non-stationarity. In order to investigate this aspect we compared, for some seizures, the correlation dimension analysis obtained for the complete seizure with that obtained for its first half.

## 3. Results

We now present the results of the correlation dimension analysis of the 19-channel EEG data for 3 patients with absences. We compare different seizures of the same patient, triggered by hyperventilation or light flashes, and seizures from different patients.

*Patient A.* Is a 16 year old boy with absences who is pharmaco-resistant. The MRI and CT scans are normal, but the SPECT shows a hipoperfusion of the right temporal and parietal lobes. The ictal EEG recording has a slightly lower amplitude on the right temporal lobe while the interictal recording exhibits a right focus also in that region.

We analysed three seizures of this patient with durations of approximately 10s, 15s and 12s, which were triggered by hyperventilation. They were recorded at 200Hz, 100Hz and 100Hz, respectively. Fig. 6 shows the map of the slopes of $\ln C_m(r)$ for one of the seizures. There one can see that part of the channels (1, 3, 4, 5, 8 and 9) exhibit a plateau, to which corresponds a correlation $D_2$, and hence seem to present chaos, whereas the other channels do not exhibit a plateau, and hence rather seem to present noise. It is important to notice that all the channels with a plateau have the same value for the correlation dimension $D_2$ and the same value for the embedding dimension m*. The analysis of the two other seizures gives similar results. In Fig. 7 we present the maps for the correlation dimension $D_2$ corresponding to the three seizures of this patient. The main feature that emerges from this analysis is that one can clearly distinguish two regions with different dynamics: the frontal and left parietal and temporal parts seem to exhibit chaotic behaviour whereas the rest rather seem to show noise. Furthermore, one finds the same pattern for the different seizures of this patient, which are also characterised by the same values of the correlation dimension $D_2$ and embedding dimension m*. For two of the seizures (the 10s and the 15s ones), we also compared the maps of $D_2$ for the complete seizures, with those for the first half of the seizures. We observed that the distinction of two dynamics regions remains in the latter (for example, the signals in Fig. 1 show the same behaviour as reported before), but some channels (located in particular in the rear part of the scalp), that did not show chaos in the former, seem now to exhibit chaos (with the $D_2$ and m$^*$ of the former). This implies that the dynamics may be changing in time, in some channels. However, because we are interested in characterising the dynamics of the seizure as a whole, we consider the analysis of the complete signal to be the proper procedure. In addition, for one of the seizures (the15s one) we analysed a bipolar montage, and found that signals associated to pairs of channels (4-10, 5-10, 5-11, 6-10, 6-11, 10-15 14-18) mainly located on the central part of the head, were giving rise to a plateau and hence a correlation dimension $D_2 \approx 1.9$.

*Patient B*. Is a 12 year old girl with absences that have responded favourably to Sodium Valproate medication.

We analysed two seizures of this patient, with duration of approximately 7s and 8s, the first being triggered by hyperventilation and the second by light flashes, they were both recorded at 100Hz. For each of the seizures one finds again two dynamic regions but now exhibiting different patterns as shown in Fig. 8, which presents the maps of the correlation dimension $D_2$ corresponding to the two seizures. The seizure triggered by hyperventilation shows a correlation dimension on the frontal and left parietal and temporal parts, in a similar way to the observed for patient A, whereas the seizure triggered by light flashes shows a left/right symmetry, with the frontal part having a correlation dimension. For both cases one finds that all the channels exhibiting a plateau have the same correlation dimension $D_2$ and embedding dimension m*.

*Patient C*. Is a 13 year old girl that has absences with walking automatism. Her epilepsy has had good evolution under Sodium Valproate medication.

We analysed two seizures of this patient, with duration of approximately 9s and 12s, triggered by hyperventilation. They were both recorded at 200Hz. We find that in both seizures none of the channels shows a plateau, that is a correlation dimension $D_2$, and hence none of the channels exhibits chaos, in contrast to those of the other two patients.

For the three patients we also analysed the EEG signals before and after the seizures, and for all the cases we did not find a plateau in any of the channels, which implies that there is no sign of chaos there.

## 4. Conclusions

The results that we obtained show that chaos analysis provides new information on the dynamics underlying absences epilepsy, and allows a distinction between situations not realised before by the more conventional forms of analysis.

In the seizures of two patients we were able to distinguish regions with different behaviour: part of the cerebral cortex seems to exhibit chaos whereas the other part seems to exhibit noise. The observation of different dynamic regions occurring in absences had not been seen before, as far as we know. The pattern is essentially the same for different seizures of the same patient, when triggered by hyperventilation, but varies for seizures of the same patient triggered either by hyperventilation or light flashes. The dynamics is characterised by a correlation dimension $D_2 \approx$ 1.7-1.9, which implies low complexity, and an embedding dimension m*$\approx$ 4-5, which indicates that the dynamics is governed by a small number of variables. The determination of the number of variables governing the dynamics is a valuable clue for model construction and possible activity control. Babloyantz and Destexhe (1986) had found $D_2 = 2.05 \pm 0.09$ and m*=5, which are in good agreement with our values. On the contrary, in the seizures of another patient we were not able to distinguish different dynamic regions and all the cerebral cortex seems to exhibit noise. Our results therefore reveal the existence of different dynamical situations associated to absences.

For the EEG signals before and after the epileptic seizures, we did not find evidence of chaotic behaviour.

It would be most desirable to understand what are the above results implying about the nature of the epileptic absences. It would also be desirable to investigate the correlation between the $D_2$ maps and the clinical picture of the patients in order to possibly use it as a diagnostic tool. In the future we would like to complement the present study with the construction of maps for the Kolmogorov entropy and the Lyapounov exponents, in order to get more information on the possible chaotic behaviour, though these studies demand longer time series and therefore require special care in the selection of the length of the seizures and the acquisition rate. Nevertheless the results that we presented here already reveal new aspects of absences epilepsy.

**Acknowledgements:**

We are grateful to F. Lopes da Silva for helpful discussions and comments.

**Figure legends:**

Fig. 1 - EEG signal recorded on patient A (seizure of 10s) at: a) channel 1, b) channel 10.

Fig. 2 - Power spectrum of the signals in Fig. 1-a) and b), respectively.

Fig. 3 - Map corresponding to the signals in Fig. 1-a) and b), respectively ( $\tau = 8 \times 10^{-2}$ s).

Fig. 4 - Correlation integral $C_m(r)$, and slope of ln $C_m(r)$, with increasing m, for the signals in Fig. 1-a) and b), respectively.

Fig. 5 – Correlation integral, $C_m(r)$, and slope of ln $C_m(r)$, with increasing m, for the control signal corresponding to the signal in Fig. 1-a).

Fig. 6 - Channel map for the slope of ln $C_m(r)$, for the seizure of 10s in patient A; the channels with a plateau have $D_2 \approx 1.7$ and $m^* \approx 5$.

Fig. 7 - Channel map for the correlation dimension $D_2$ of three seizures from patient A, triggered by hyperventilation, with duration: a)10s, b)15s and c) 12s; for all of them $D_2 \approx 1.7$, $m^* \approx 5$ for a) and $m^* \approx 4$ for b) and c).

Fig. 8 - Channel map for the correlation dimension $D_2$ of two seizures from patient B, triggered by: a) hyperventilation, and b) light flashes, with duration 7s and 8s, respectively; for both $D_2 \approx 1.9$ and $m^* \approx 4$.